\documentclass{raa_twocolumn}

\usepackage{graphicx,times}             
\usepackage{grffile}
\usepackage{natbib}
\usepackage{amssymb,amsmath}
\usepackage{multirow}
\usepackage{multicol}
\usepackage{CJK}
\usepackage{epstopdf}
\usepackage{caption}
\usepackage{subcaption}
\usepackage{wasysym}
\usepackage{natbib}
\bibpunct{(}{)}{;}{a}{}{,}

\usepackage[pagebackref=true]{hyperref}
\hypersetup{colorlinks = true, linkcolor = green, anchorcolor = red, citecolor = blue, filecolor = red, pagecolor = red, urlcolor = red}

\begin{document}
\begin{CJK*}{UTF8}{gbsn}
  \title{An FUor-like Outbursting Class I Protostar in NGC~7538}

   \volnopage{Vol.0 (20xx) No.0, 000--000}      
   \setcounter{page}{1}          

   \author{Aidi Fang (方艾迪)  
      \inst{1,2}\thanks{Based on observations obtained with WIRCam, a joint project of CFHT, Taiwan, Korea, Canada, France, at the Canada-France-Hawaii Telescope (CFHT) which is operated by the National Research Council (NRC) of Canada, the Institute National des Sciences de l'Univers of the Centre National de la Recherche Scientifique of France, and the University of Hawaii.}
   \and Zhiwei Chen (陈志维) 
      \inst{1,2}\thanks{Corresponding author: chenzhiwei@ctgu.edu.cn}
      \and Lin Du (杜林)
      \inst{3,4}
      \and Sheng Zheng (郑胜)
      \inst{1,2}
   }
  \institute{
  College of Science, China Three Gorges University, Yichang 443002, People's Republic of China
  \and
   Center for Astronomy and Space Sciences, China Three Gorges University, Yichang 443002, People's Republic of China
   \and
   Key Laboratory of Optical Astronomy, National Astronomical Observatories, Chinese Academy of Sciences, 20A Datun Road, Chaoyang District, Beijing 100101, People's Republic of China
   \and
   School of Astronomy and Space Science, University of Chinese Academy of Sciences, Beijing 100049, People's Republic of China
\vs\no
   {\small Received 2026 Feb 09; accepted 2026 Apr 03}}

\abstract{
We report on the discovery of an FUor-like Class I protostar in NGC~7538. The source, named NGC~7538~MIR, exhibited a giant luminosity burst ($\Delta K_s\sim5$) and a prolonged high-luminosity state lasting at least five years. Its mid-infrared (mid-IR) light curves, constructed from WISE/NEOWISE multiepoch data, presented a rapid rise and slight fading after the peak, placing this event among long-duration eruptive phenomena observed in protostars, for example, FUor-type events. The evolution of W1/W2 luminosity and $W1-W2$ color can be naturally split into three phases, pre-burst, burst and post-burst, suggesting that different physical processes may dominate in the three phases. The evolution of NGC~7538~MIR is consistent with a transition from variability influenced by circumstellar extinction (pre-burst) to a phase with greatly enhanced accretion luminosity (burst), and followed by a gradual relaxation of the circumstellar environment (post-burst). Overall, the observed IR variability of NGC~7538~MIR is consistent with an FUor-like accretion event occurred at an early evolutionary stage, highlighting the importance of long-term IR monitoring for identifying episodic accretion events in deeply embedded protostars.
}


   \maketitle

%
%
\section{Introduction}           
\label{sect:intro}

Young stellar objects (YSOs) exhibit photometric variability over a broad range of timescales and amplitudes. Although low-amplitude variations can arise from stellar rotation, surface inhomogeneities, or variable circumstellar extinction, large-amplitude luminosity changes are generally attributed to temporal variations in the mass accretion rate from the circumstellar disk onto the central protostar \citep{1977ApJ...217..693H,1996ARA&A..34..207H}. Episodic accretion is now widely recognized as a fundamental process during early stellar evolution, contributing substantially to stellar mass assembly during embedded phases and exerting a long-term influence on disk evolution and planet formation \citep{1997ApJ...475..770H}.

Among eruptive accretion phenomena, FUors represent the most extreme manifestations, characterized by sudden increases in luminosity of several magnitudes followed by elevated accretion states that can persist for decades or longer \citep{1977ApJ...217..693H}. Classical FUors were initially identified through optical observations and subsequently interpreted as episodes of dramatically enhanced disk accretion, during which the accretion disk dominates the emergent spectrum \citep{1996ARA&A..34..207H}. Traditionally, eruptive YSOs have been classified into FUors and shorter-duration, recurrent EX Lupi-type objects \citep{1989ESOC...33..233H}. However, recent wide-field infrared surveys have revealed a growing population of eruptive sources whose amplitudes, durations, and evolutionary stages span a continuous range of properties, challenging the classical FUor-EXor dichotomy \citep{Audard_2014}.

In this context, long-term IR photometric monitoring has emerged as a powerful tool for identifying FUor-like accretion outbursts. Objects classified as FUor-like exhibit photometric and spectroscopic properties consistent with FUors, even when the initial outburst was not observationally recorded \citep{2003AJ....126.2936A,2018ApJ...861..145C}. Large-amplitude IR brightening accompanied by systematic color evolution provides strong evidence for enhanced disk accretion, particularly for deeply embedded Class~I sources where optical spectroscopy is often infeasible \citep{2025JKAS...58..209C}. Such photometric diagnostics have proven essential for extending the census of FUor-like objects beyond the classical, optically visible population.

NGC 7538 is a prominent massive star-forming complex located in the Perseus spiral arm at a distance of approximately 2.65 kpc \citep{2009ApJ...693..406M}. It corresponds to the optical H II region Sh2-158 and is associated with the Cas OB2 complex \citep{1959ApJS....4..257S,2008ApJ...685L..51F}. The region hosts a series of luminous IR sources, commonly designated IRS 1 through IRS 11, which mark multiple sites of high-mass star formation at different evolutionary stages. These IRS sources, including compact and ultra-compact H II regions, deeply embedded massive protostars, molecular outflows, and dense dust condensations, have been extensively investigated in previous studies \citep[e.g.,][IRS~1-11]{Puga_2010,2014A&A...567A.116F,10.1093/mnras/stu1396/2014,2020ApJ...904..139S,2023MNRAS.519.1013B}. More recent multi-wavelength observations provide an updated view of the stellar content and cloud structure of NGC~7538. These surveys have identified a large population of YSO candidates, predominantly Class I and Class II objects, confirming ongoing and spatially structured star formation across the molecular cloud \citep{Chavarr_a_2014,10.1093/mnras/stu1396/2014,2017MNRAS.467.2943S}.

\begin{figure*}[ht]
\centering
\includegraphics[width=0.8\linewidth]{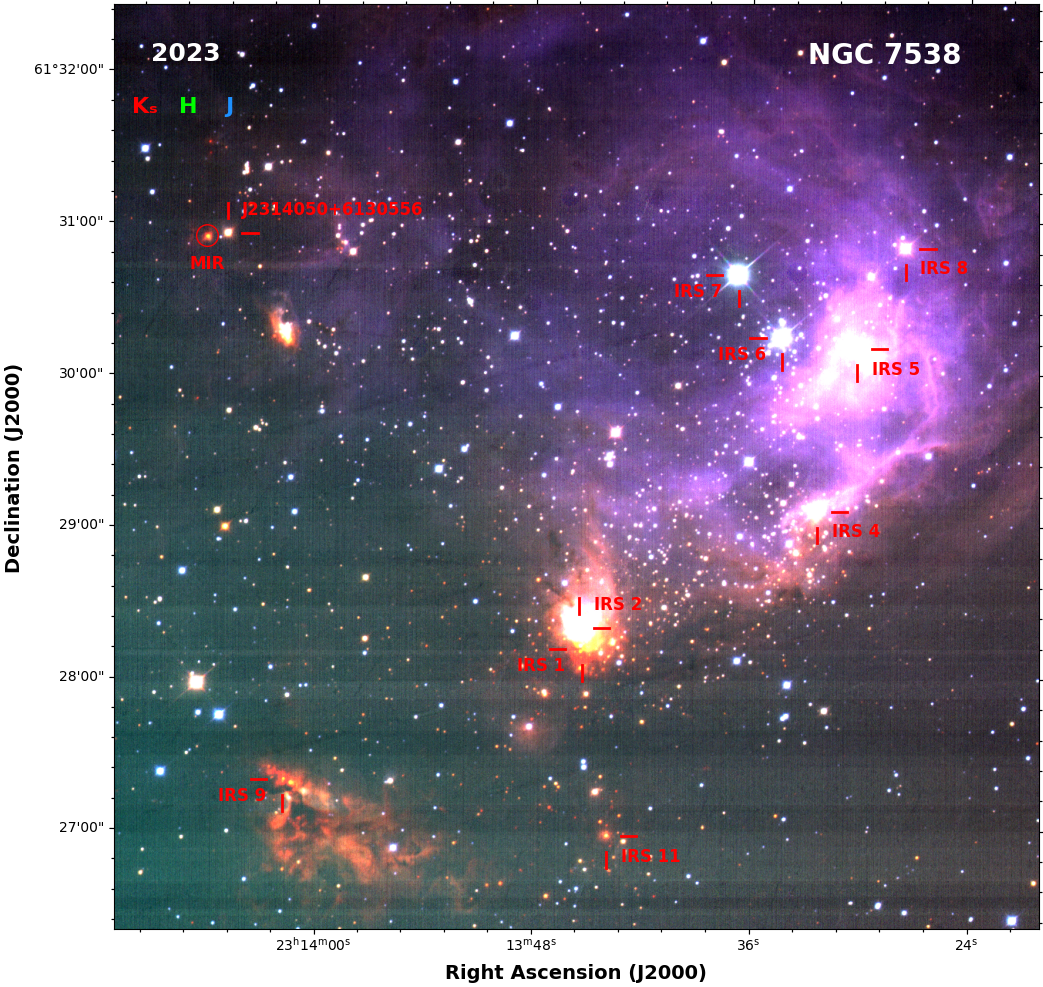}
\caption{Three-color image of NGC~7538 (blue: $J$, green: $H$, red: $K_s$), made from the CFHT observations in 2023 September.}
\label{Fig1}
\end{figure*}

The southern and western edges of NGC 7538, where IR-bright YSOs and submillimeter clumps are located \citep[e.g.,][]{2005ApJ...625..891R}, have received intense interest over the past several decades. However, the northeast of NGC 7538 has been much less explored. In this paper, we report on the discovery and new near-IR observations for an FUor-like outbursting Class I protostar NGC~7538~MIR in this area. Figure~\ref{Fig1} shows the $JHK_s$ three-color image of NGC 7538 obtained in 2023 September. This source is near the northeastern edge of the H~{\sc ii} region, but has received relatively little observational attention. \citet{Chavarr_a_2014} classified this source as a Class I protostar on the basis of \textit{Spitzer} IRAC observations in the $3-8\mu$m wavelength range and near-IR $H$ and $K$ observations. In particular, NGC~7538~MIR coincides with the massive dense clump SMM 63(A) detected by submillimeter observations at $450$ and $850\,\mu$m with the JCMT \citep{2005ApJ...625..891R}. The integrated fluxes of SMM 63(A) at $450$ and $850\,\mu$m suggest a clump mass of $\sim140\,M_\odot$ for a dust temperature of 35 K \citep{2005ApJ...625..891R}. The association between NGC~7538~MIR and SMM 63(A) strongly supports the interpretation that NGC~7538~MIR is most likely a deeply embedded Class I protostar.

Given the large distance and likely large extinction to NGC~7538~MIR, the observed variability amplitude ($\Delta K_s\sim 5$) implies a substantial outburst in accretion luminosity, suggesting that the object may lie in the intermediate- to high-mass regime. Although accretion-driven outbursts and photometric variability have been extensively studied in low-mass YSOs, similar phenomena in intermediate- and high-mass YSOs remain observationally rare and poorly constrained \citep{2021ApJ...922...90C,2025AJ....170..125C}. As such, NGC~7538~MIR offers an important opportunity to investigate episodic accretion and energetic variability in eruptive YSOs of higher mass.

In this paper, we present the discovery and IR characterization of a large-amplitude eruptive event in the deeply embedded Class~I protostar NGC~7538~MIR. By combining multi-epoch near- and mid-IR photometric data from CFHT and WISE/NEOWISE, we construct a long-term light curve spanning more than a decade and investigate the luminosity and color evolution of the source. Particular emphasis is placed on using IR color-magnitude diagnostics to disentangle the roles of accretion-driven luminosity enhancement and variable circumstellar extinction in the absence of optical constraints. The observations and data reduction are described in Sect.~\ref{sect:obs}. The results are presented in Sect.~\ref{sect:result}. In Sect.~\ref{sect:analysis}, we discuss the physical mechanisms responsible for the observed variability and the nature of NGC~7538~MIR. The main conclusions are summarized in Sect.~\ref{sect:conclusion}.


\section{Multiepoch Infrared Observations and Data Reduction}
\label{sect:obs}

\subsection{Canada-France-Hawaii Telescope (CFHT) near-IR Observations}
The Wide-field InfraRed Camera (WIRCam; \citealt{2004SPIE.5492..978P}) is a near-IR wide-field imaging instrument mounted at CFHT. WIRCam consists of a mosaic of four HAWAII-2RG detectors, each with $2048 \times 2048$ pixels, providing a pixel scale of $0\farcs3\,\mathrm{pixel^{-1}}$ and a total field of view of $20\arcmin \times 20\arcmin$. Near-IR observations for NGC 7538 in the $JHK_s$ bands were made in 2023 September as part of CFHT observing program 23BS15. In order to track long-term IR variability, we searched for archival CFHT-WIRCam observations of NGC 7538 from Canadian Astronomy Data Centre (CADC) \footnote{https://www.cadc-ccda.hia-iha.nrc-cnrc.gc.ca/en/cfht}. The CFHT-WIRCam observations in 2006 July, 2011 September, and 2012 August were retrieved. 

All the CFHT-WIRCam observations in four epochs were reduced with the I'iwi pipeline, the official data processing and calibration pipeline for WIRCam. The processed and calibrated images of NGC 7538 in four epochs were obtained from CADC.

\subsection{Near-IR Photometry for CFHT-WIRCam Images}

Aperture photometry was performed for stacked $JHK_s$ images in four epochs using the \textit{photutils} package \citep{Bradley2025}, a coordinated package of \textit{Astropy} \citep{2013A&A...558A..33A,2018AJ....156..123A}. Circular apertures with radii scaled to the measured full width at half maximum (FWHM) of point sources were adopted. To determine the optimal aperture size, ten aperture radii were applied uniformly sampling the range from 1.5 to 3.5 times FWHM, and aperture photometry was performed independently for each aperture with \textit{photutils}. The final aperture was selected on the basis of the convergence of the growth curve by choosing the radius at which the enclosed flux becomes stable against further increases in aperture size. The local sky background for each source was estimated and subtracted using concentric annular apertures with an inner radius of 4 times FWHM and a width of 1.5 times FWHM, effectively minimizing contamination from the source flux while accounting for spatial variations in the IR background. Instrumental magnitudes were calibrated using unsaturated stars from the 2MASS point source catalog \citep{2006AJ....131.1163S}, yielding the final calibrated photometry in the $JHK_s$ bands.

\subsection{WISE/NEOWISE Mid-IR Data}
The Wide-field Infrared Survey Explorer (WISE) \citep{2010AJ....140.1868W} is a 40\,cm space telescope in a low-Earth orbit that surveyed the entire sky in 2010 using four infrared bands at 3.4, 4.6, 12, and 22\,$\mu$m. The angular resolutions in the four bands ($W1$, $W2$, $W3$, and $W4$) are $6\farcs1$, $6\farcs4$, $6\farcs5$, and $12\farcs0$, respectively. With the primary aim of studying near-Earth objects, NEOWISE observations resumed in 2013 December and have continued to the present, following the depletion of the telescope's cryogenic coolant. Due to the orbit of the telescope, WISE/NEOWISE observations visit NGC 7538 twice a year, in January and July. From the NEOWISE 2024 data release containing $W1$ and $W2$ observations from 2014 January to 2024 July, the single-exposure data of NGC 7538 in $W1$ and $W2$ bands collected over 11 years (22 epochs) are analyzed in this work, in addition to the WISE all-sky survey images of NGC~7538 in January and July 2010.

Although time-domain photometry is available from unTimely: a Full-sky, Time-Domain unWISE Catalog \citep{Meisner_2023}, the cataloged light curves of NGC~7538~MIR are incomplete and insufficient for a reliable characterization of its variability. In particular, during the outburst phase, the pronounced increase in flux combined with the limited angular resolution of the WISE/NEOWISE instruments results in significant blending between the target and nearby source J2314050+6130556, preventing a clear separation between them.

\begin{figure*}
\centering
\includegraphics[width=\textwidth, angle=0]{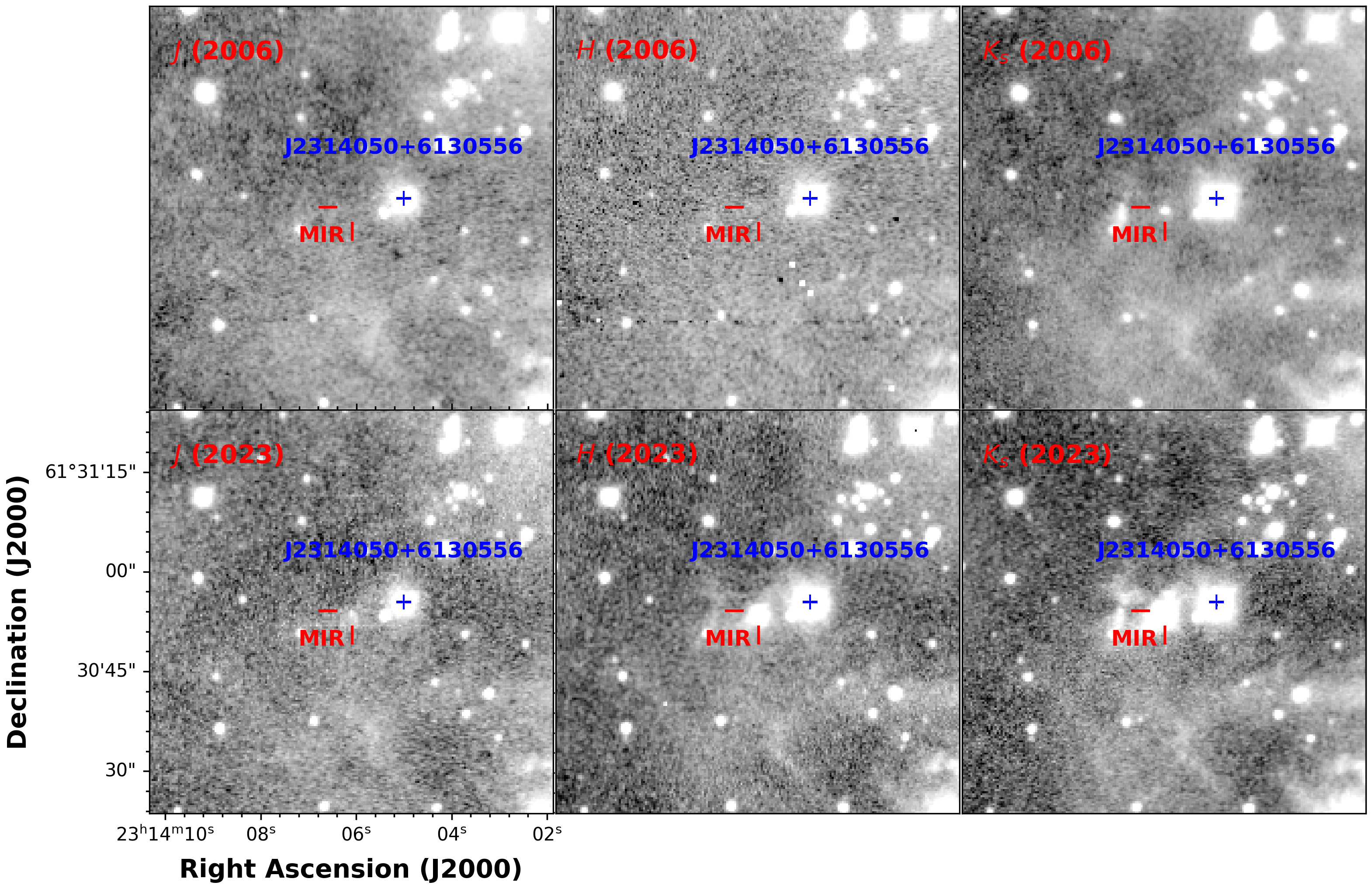}
\caption{$JHK_s$ images of NGC~7538~MIR in two epochs (2006 and 2023).
}
\label{Fig2}
\end{figure*}

\begin{table*}
\centering
\caption{Near-IR $JHK_s$ photometry of NGC 7538 MIR}
\begin{tabular}{ccccccc}
\hline
Index & Date & MJD & $J$ & $H$ & $K_s$ & Telescope \\
& & & (mag) & (mag) & (mag) &  \\
\hline
1 & 2004-12-01 & 53348.0 & ... & 18.585$\pm$0.066 & 15.690$\pm$0.065 & KPNO\textsuperscript{a}\\
2 & 2006-07-09 & 53925.0 & ... & ... & 15.828$\pm$0.044 & CFHT \\
3 & 2011-09-18 & 55820.0 & ... & ... & 15.164$\pm$0.041 & CFHT \\
4 & 2012-08-29 & 56167.0 & ... & ... & 14.965$\pm$0.042 & CFHT \\
5 & 2023-09-02 & 60193.0 & 17.271$\pm$0.053 & 14.042$\pm$0.041 & 11.165$\pm$0.029 & CFHT \\
\hline
\end{tabular}
  \label{tab:NIR}
  \\
\textsuperscript{a}{Photometry adopted from \cite{Chavarr_a_2014}.}
\end{table*}

We therefore performed photometric measurements directly on the $W1$ and $W2$ images of WISE/NEOWISE. The photometric procedure adopted for the WISE/NEOWISE data is identical to that applied to the CFHT images, employing adaptive aperture photometry based on the measured FWHM of point sources, curve-of-growth convergence to determine the optimal aperture, and local sky subtraction using round annulus. The photometry in the $W1$ and $W2$ bands was tied to the unTimely catalog on the basis of isolated sources. 

In order to extract the $W1$ and $W2$ fluxes of NGC 7538 MIR in several epochs, the photometry procedure is different from other isolated sources. We first measured the total flux of NGC 7538 MIR and J2314050+6130556 with a fixed aperture for all epochs in $W1$ and $W2$ bands. The aperture is centered between the two sources and has a radius large enough to encompass the two sources. The nearby source J2314050+6130556 is assumed to be stable in all epochs. In the epoch of 2010 January, NGC 7538 MIR is much fainter than J2314050+6130556 in $W1$ and $W2$ bands. The $W1$ and $W2$ photometry in this epoch was obtained for J2314050+6130556, and was checked with the unTimely and ALLWISE catalogs \citep{AllWISECatalog}. The $W1$ and $W2$ fluxes of J2314050+6130556 are assumed to be stable in all epochs. Thus, the $W1$ and $W2$ photometry of NGC 7538 MIR in every epoch was obtained by subtracting the fluxes of J2314050+6130556 from the total fluxes of the two sources. We obtained the $W1$ and $W2$ fluxes of NGC~7538~MIR in 24 epochs from 2010 January to 2024 July.

\begin{figure*}
\centering
\includegraphics[width=\textwidth, angle=0]{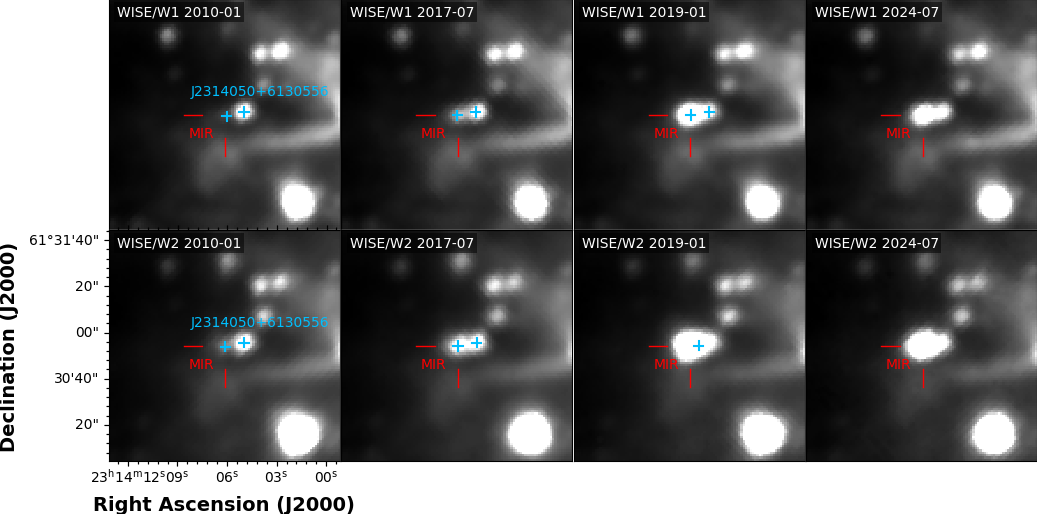}
\caption{WISE/NEOWISE W1 (top) and W2 (bottom) images at four representative epochs (2010, 2017, 2019, and 2024). The target position is marked by red crosses; blue crosses denote catalog positions from the untimely catalog \citep{Meisner_2023}.
}
\label{Fig3}
\end{figure*}

\section{Result}
\label{sect:result}
\subsection{Near-IR Variability}

The outburst source NGC~7538~MIR is located on the northeast side of the NGC~7538 \ion{H}{ii} region (see Figure~\ref{Fig1}). Its position, measured from the CFHT $H$-band image obtained in September 2023, is $\alpha$(J2000) = 23$^{\rm h}$14$^{\rm m}$06.13$^{\rm s}$, $\delta$(J2000) = +61$^\circ$30$^\prime$54.12$^{\prime\prime}$, consistent with $HK_s$-band detection in \cite{Chavarr_a_2014}.

Figure~\ref{Fig2} depicts the near-IR variability of NGC~7538~MIR by comparing CFHT $JHK_s$ images in two epochs (2006 and 2023). NGC~7538~MIR was invisible in the $JH$ bands and faint in the $K_s$ band in 2006. Its near-IR luminosity increased dramatically in recent CFHT $JHKs$ images in 2023. The $K_s$-band image in 2023 indicates nebulosity in the close vicinity of NGC~7538~MIR, similar to other outburst YSOs with jet activities \citep[e.g.,][]{2017NatPh..13..276C}. In contrast, the nearby 2MASS source J2314050+6130556 is stable in the $JHK_s$ bands over the same time span. This source maintained a mean magnitude of approximately 10.757 with variance of about 0.002 across the four observational epochs.

Table~\ref{tab:NIR} reports near-IR $JHK_s$ photometry of NGC~7538~MIR in five epochs from 2004 to 2023. The $K_s$-band luminosity of NGC~7538~MIR continuously increased in this period. The highest magnitude of variability reaches $\Delta K_s\sim4.7$ for the epochs in 2006 and 2023. It should be noted that the near-IR luminosity of the source may not have increased monotonically between 2004 and 2023. While the data suggest a possible luminosity minimum around 2006, the observed brightness in 2023 cannot be definitively established as the peak value over this period. Indeed, the WISE/NEOWISE variability of NGC~7538~MIR peaks between 2018 and 2019 (see Section~\ref{sect:mir_var} ). The CFHT observations in 2023 are in the post-burst phase, implying a slightly brighter magnitude $K_s\sim11$ in $2018-2019$. Therefore, NGC~7538~MIR is likely to have an IR outburst magnitude of $\Delta K_s\simeq5$, similar to those of FUor-type YSOs \citep[e.g.,][]{Audard_2014,Fischer2023,2025JKAS...58..209C}.

\begin{table*}
\centering
\caption{WISE/NEOWISE Photometry of NGC 7538 MIR}
\label{tab:photometry}
\begin{tabular}{ccccc}
\hline
Index & Date & MJD & $W1$ & $W2$  \\
      &      &     & (mag) & (mag)   \\
\hline
1 & 2006-08-14 & 53961.0 & 12.186$\pm$0.029 & 10.606$\pm$0.029\textsuperscript{a} \\
2 & 2010-01-16 & 55211.0 & 12.052$\pm$0.063 & 10.413$\pm$0.038  \\
3 & 2010-07-23 & 55399.0 & 11.863$\pm$0.060 & 10.311$\pm$0.034  \\
4 & 2014-01-28 & 56674.0 & 11.820$\pm$0.057 & 10.314$\pm$0.033 \\
5 & 2014-08-05 & 56863.0 & 11.781$\pm$0.056 & 10.309$\pm$0.032   \\
6 & 2015-01-26 & 57037.0 & 11.689$\pm$0.054 & 10.182$\pm$0.030   \\
7 & 2015-07-23 & 57225.0 & 11.582$\pm$0.053 & 10.133$\pm$0.031  \\
8 & 2016-01-12 & 57398.0 & 11.291$\pm$0.052 & 9.879$\pm$0.031   \\
9 & 2016-07-22 & 57590.0 & 11.122$\pm$0.047 & 9.673$\pm$0.026   \\
10 & 2017-01-05 & 57758.0 & 11.021$\pm$0.034 & 9.544$\pm$0.023  \\
11 & 2017-07-22 & 57956.0 & 10.873$\pm$0.025 & 9.431$\pm$0.023  \\
12 & 2018-01-01 & 58120.0 & 9.894$\pm$0.024 & 8.372$\pm$0.024   \\
13 & 2018-07-20 & 58321.0 & 8.934$\pm$0.023 & 7.210$\pm$0.025  \\
14 & 2019-01-05 & 58481.0 & 8.912$\pm$0.024 & 7.224$\pm$0.026   \\
15 & 2019-07-19 & 58686.0 & 9.041$\pm$0.026 & 7.443$\pm$0.027  \\
16 & 2020-01-04 & 58846.0 & 9.148$\pm$0.029 & 7.591$\pm$0.025 \\
17 & 2020-07-18 & 59052.0 & 9.167$\pm$0.028 & 7.569$\pm$0.025   \\
18 & 2021-01-03 & 59213.0 & 9.161$\pm$0.028 & 7.532$\pm$0.025  \\
19 & 2021-07-16 & 59416.0 & 9.173$\pm$0.028 & 7.512$\pm$0.024   \\
20 & 2022-01-01 & 59577.0 & 9.169$\pm$0.027 & 7.510$\pm$0.025  \\
21 & 2022-07-15 & 59782.0 & 9.181$\pm$0.027 & 7.583$\pm$0.026   \\
22 & 2023-01-05 & 59941.0 & 9.184$\pm$0.028 & 7.522$\pm$0.025 \\
23 & 2023-07-20 & 60147.0 & 9.201$\pm$0.029 & 7.574$\pm$0.025  \\
24 & 2024-01-05 & 60308.0 & 9.234$\pm$0.029 & 7.571$\pm$0.025  \\
25 & 2024-07-18 & 60512.0 & 9.281$\pm$0.030 & 7.589$\pm$0.026  \\
\hline
\end{tabular}
\\
\textsuperscript{a}{\textit{Spitzer} photometry at $3.6$ and $4.5\,\mu$m adopted from \cite{Chavarr_a_2014}.}

\end{table*}

\subsection{Mid-IR Variability}\label{sect:mir_var}
The near-IR observations for NGC~7538~MIR are insufficient to better constrain the variable nature of the source. The WISE/NEOWISE data in 24 epochs from 2010 to 2024, and the Spitzer data in 2006 \citep{Chavarr_a_2014}, can provide detailed IR variability, which is crucial to characterizing the variability type for this Class I protostar in the massive star-forming region NGC~7538.

Figure~\ref{Fig3} presents the WISE/NEOWISE W1 and W2 images of NGC~7538~MIR in four epochs from 2010 to 2024. NGC~7538~MIR was much fainter than the nearby source J2314050+6130556 in the first epoch 2010. Seven and half years later, in 2017, NGC~7538~MIR was comparable to J2314050+6130556. NGC~7538~MIR reached its maximum luminosity in 2019 and remains in a high state in 2024. The $W1$ and $W2$ photometry of NGC~7538~MIR in 25 epochs from 2006 to 2024 is listed in Table~\ref{tab:photometry}. NGC~7538~MIR was in a low state before 2010, followed by a gradual brightening in $W1$ and $W2$ from July 2010 to July 2017. Afterwards, NGC~7538~MIR reached its maximum luminosity in 1 year, and remained in a high state but with a slow trend in fading from 2019 July to 2024 July. The largest amplitude of variability is $\Delta W1\sim3.1$ and $\Delta W1\sim3.2$, much smaller than the $K_s$-band amplitude ($\Delta K_s\sim5$).

\subsection{Light curves of Infrared Luminosities and Colors}

\begin{figure*}
\centering
\includegraphics[width=\textwidth, angle=0]{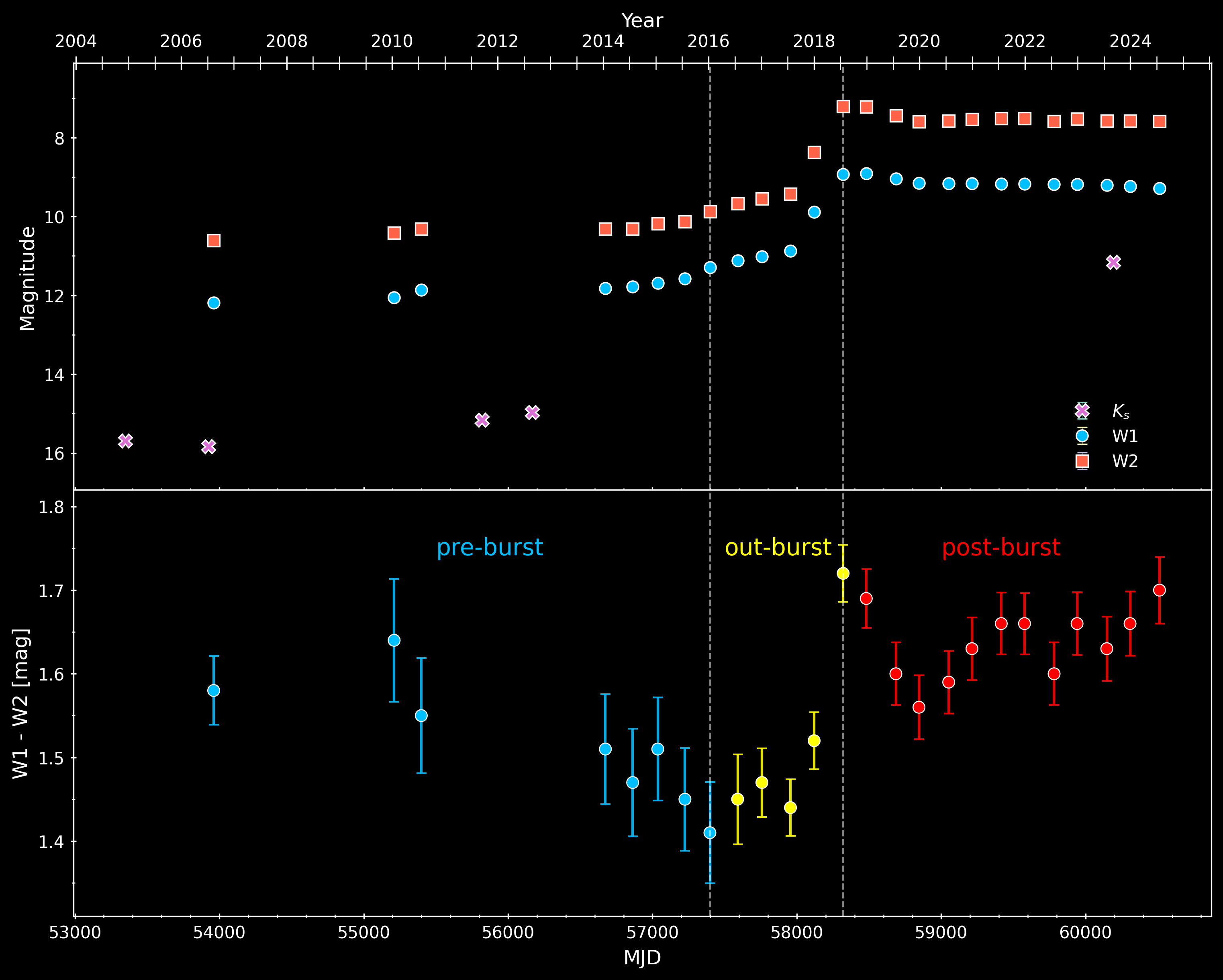}
\caption{Multi-wavelength light curve and color evolution of NGC~7538~MIR. (top) Light curves in the $K_s$, $W1$, and $W2$ bands, showing a transition from quiescence to a sustained high-luminosity state. (bottom) Evolution of the $W1-W2$ color index, illustrating distinct photometric phases during the outburst.}
\label{Fig4}
\end{figure*}

The long-term near-IR and mid-IR light curves of NGC~7538~MIR reveal a large-amplitude outburst spanning more than a decade (see the top panel of Figure~\ref{Fig4}). The mid-IR light curve with a cadence of 6 months reveals that NGC~7538~MIR reached its maximum luminosity between 2018 July and 2019 January, with IR outbursts of $\Delta K_s\sim5$, $\Delta W1\sim3.1$ and $\Delta W2\sim3.2$. Following the peak luminosity in 2019, the source entered a prolonged high-state lasting more than five years, with only moderate fading in $W1$ and $W2$.

\begin{figure*}
\centering
\includegraphics[width=\textwidth, angle=0]{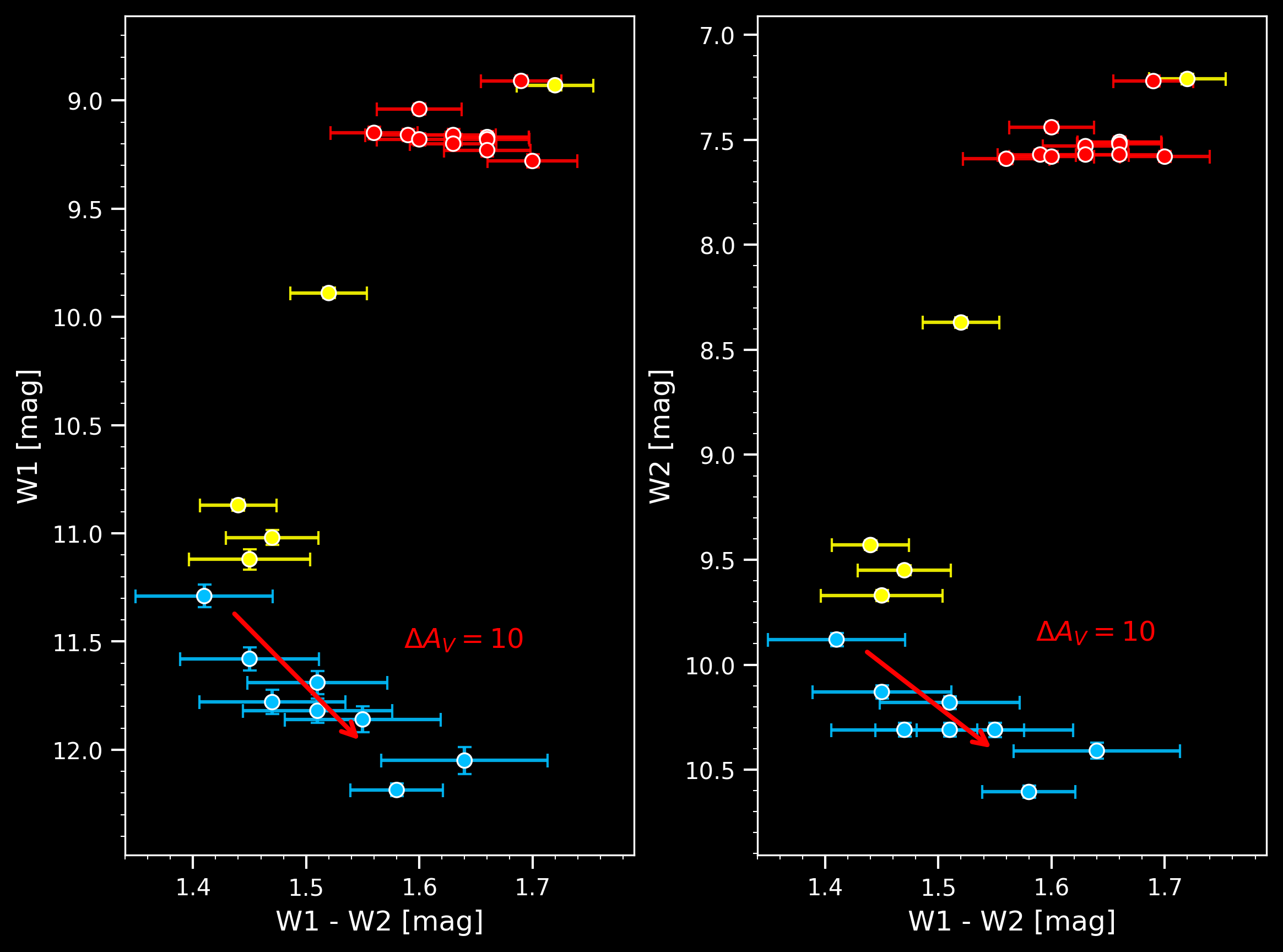}
\caption{Magnitude versus color diagrams for $W1$ and $W2$ bands. Data points are color-coded by epoch: blue (2006 - 2016), yellow (2016 - 2018) and red (2019 - 2024), corresponding to different variability phases of the source. The solid line indicates the extinction vector in the mid-IR, calculated following the extinction laws of \cite{Yuan_2013} and \cite{Zhang_2022}. }
\label{Fig5}
\end{figure*}

In addition to the pronounced variability in the near-IR and mid-IR, NGC~7538~MIR exhibits significant $W1 - W2$ color changes throughout the eruptive event. The bottom panel of Figure~\ref{Fig4} shows the variation of $W1 - W2$ color with time. The evolution of $W1/W2$ luminosity and $W1-W2$ color can be clearly split into three phases: pre-burst phase before 2016, outburst phase between 2016 and 2019, and post-burst phase from 2019 to 2024. The three phases are folded with different colors, as shown in Figure~\ref{Fig4}. In Figure~\ref{Fig5}, the time-dependent evolution of NGC~7538~MIR in the magnitude versus $W1-W2$ color diagrams is folded in the same colors as in Figure~\ref{Fig4}. The three phases of the source are naturally separated in Figure~\ref{Fig5}, implying that the three phases are intrinsically different in their physical states.

\section{Discussion}
\label{sect:analysis}

\subsection{Physical Mechanisms of the Multi-phase Evolution}

The outbursting source underwent three phases, during which the source presented distinct properties in luminosity and color evolution. In the following, we discuss the time-dependent evolution along the three phases and the physical mechanisms of the multi-phase evolution.\\

Pre-burst phase (before 2016). During this phase, the source exhibits a slow and monotonic brightening in all infrared bands. Quantitatively, the source brightened by $\sim$0.9 mag in $W1$ and $\sim$0.7 mag in $W2$, accompanied by a bluing change of $\Delta(W1-W2) = -0.17$. The $W1-W2$ color shows a clear bluer-when-brighter (BWB) trend \citep{2025ApJS..278...10N}, and the corresponding trajectory in the magnitude--color diagram is parallel to the reddening vector (Figure~\ref{Fig5}), consistent with extinction-driven variability reported in other eruptive YSOs \citep[e.g.,][]{2014AJ....148...11K,2019AJ....158..240H,2021ApJ...923..171P,Morris2025MNRAS}.

The BWB behavior observed in the mid-infrared can be naturally explained by variations in circumstellar extinction. Structural changes in the inner disk or inhomogeneous dust distributions can reduce the line-of-sight obscuration toward the protostar and the innermost disk regions. Since the $W1$ band traces hotter emission and is more sensitive to extinction than $W2$, a decrease in extinction produces a larger brightening at shorter wavelengths, resulting in a BWB trend as the source brightens \citep{1989ESOC...33..233H, Parks_2014}. This interpretation is further supported by the Class~I nature of the source \citep{Chavarr_a_2014}. Class~I YSOs are embedded in dense, highly structured envelopes and statistically exhibit a higher incidence of mid-IR BWB behavior than more evolved systems, consistent with extinction-driven variability dominating at early evolutionary stages \citep{2021ApJ...923..171P}.\\

Outburst phase (2016--2019). During this phase, the source undergoes a rapid and monotonic brightening in all IR bands, reaching its maximum luminosity between 2018 July and 2019 January. The variability amplitudes are $\sim$2.2 mag in $W1$ and $\sim$2.5 mag in $W2$, accompanied by a reddening change of $\Delta(W1-W2) = 0.27$. The color evolution shows a clear RWB trend \citep{2025ApJS..278...10N}, and the trajectory in the magnitude--color diagram deviates significantly from the reddening vector, indicating that extinction alone cannot account for the observed variability \citep[e.g.,][]{2020ApJ...889..148K,2021ApJ...920..132P,Morris2025MNRAS}.

As the source approaches the outburst peak, it exhibits a clear redder-when-brighter (RWB) trend, indicative of accretion-dominated variability. A rapid increase in the disk accretion rate leads to enhanced heating of the inner disk and surface layers, substantially raising the accretion luminosity. This excess energy is absorbed and reprocessed by circumstellar dust and re-emitted at IR wavelengths, causing mid-IR emission to be increasingly dominated by warm dust in the inner disk and envelope \citep{Hillenbrand_2013, Audard_2014}. Because the $W2$ band is more sensitive to cooler dust emission than $W1$, enhanced reprocessed emission results in relatively stronger brightening at longer wavelengths, naturally producing an RWB trend. This behavior near outburst peaks has been widely reported in eruptive YSOs and is commonly interpreted as a signature of intrinsic luminosity increases driven by enhanced accretion, rather than extinction variations \citep{Hillenbrand_2013, 2017MmSAI..88..777C, 2021ApJ...923..171P}. The clear deviation from the extinction vector during this phase therefore supports an accretion-driven origin for the observed mid-IR variability.\\

Post-outburst phase (2019--2024). Following the luminosity peak, the source enters a prolonged high-luminosity state lasting more than five years, with only moderate fading ($\Delta W1 \sim -0.2$ mag and $\Delta W2 \sim -0.4$ mag). The $W1-W2$ color exhibits a more complex evolution, showing an initial bluing ($\Delta(W1-W2) = -0.13$) followed by a mild reddening ($\Delta(W1-W2) = 0.08$). The trajectory in the magnitude--color diagram displays a cyclic pattern, which cannot be explained by a simple extinction vector \citep{2021ApJ...920..132P}. Similar cyclic color variations have also been reported in other YSOs \citep[e.g.,][]{2019AJ....158..240H,2020ApJ...903....5L,2024MNRAS.52711651A}.

During this phase, the source exhibits a counterclockwise color variation, which can be consistently interpreted as extinction variations driven by the evolution of the $\sim$3.05~$\mu$m H$_2$O ice absorption feature. Near the outburst peak, enhanced accretion luminosity heats the inner envelope and disk surface layers, leading to sublimation of ice mantles on silicate dust grains and a significant weakening of the water-ice feature \citep{1998ApJ...492..213G, Brooke_1999, Shimonishi_2008}. Because this absorption preferentially suppresses flux in the W1 band, its removal results in a relative enhancement of $W1$ compared to $W2$, producing bluer mid-IR colors without requiring an increase in intrinsic temperature. As the accretion luminosity continues to decline, water ice gradually re-condenses onto dust grains, re-establishing the ice absorption feature and leading to a mild reddening trend at late times \citep{2002MNRAS.336..425H}. Such delayed recovery of ice features has been observed in embedded YSOs and is supported by recent JWST spectroscopy tracing ice evolution over 1.7--5.3~$\mu$m \citep{2025ApJ...992..203D}. The combined fading and color evolution therefore indicate that the late-time mid-IR variability is governed by the thermal and chemical relaxation of the circumstellar environment, reflecting the competitive interplay between declining accretion luminosity and extinction variations \citep{2020ApJ...903....5L}.

\subsection{NGC~7538~MIR, a new FUor-like accretion event?}
\label{sect:future}

The infrared outburst properties of NGC~7538~MIR are indicative of an FUor-like accretion event rather than a V1647~Ori-type eruption. While both categories are manifestations of enhanced disk accretion with overlapping photometric amplitudes, they are distinguished by key observational signatures and evolutionary timescales. V1647~Ori-type events occupy an intermediate phenomenological space, with outburst durations longer than typical EX Lup bursts (months) but shorter than classical FUors (decades to centuries) \citep{2017MmSAI..88..777C}. Their spectral characteristics often represent a hybrid, displaying features of both high-state EX Lup objects (e.g., emission lines like Pa$\beta$) and FUors (e.g., water absorption, weak metals) \citep{2004ApJ...606L.123B, 2007AJ....133.2020A}. In contrast, classical FUor outbursts are defined by a transition to a prolonged, high-accretion state where a viscously heated disk dominates the luminosity output, often for human-lifetime scales. This results in a sustained luminosity plateau, distinctive wavelength-dependent absorption spectra, and line profiles consistent with disk rotation \citep{1996ARA&A..34..207H, 2018ApJ...861..145C}. The $>5$ year persistence of NGC~7538~MIR near its peak IR brightness, with only modest fading, aligns with this long-lived high-state characteristic and is difficult to reconcile with the shorter decay timescale expected for V1647~Ori-type events.

The deeply embedded, Class~I nature of NGC~7538~MIR also explains the absence of classical optical spectroscopic diagnostics typically associated with FUors. It is now established that FUor-type accretion eruptions can occur in early-stage protostars (Class~0/I), where the outburst physics is primarily probed through infrared and sub-millimeter luminosity evolution rather than optical spectroscopy \citep{Safron_2015, 2019ApJ...872..183F}. Spectroscopic observations of NGC~7538~MIR, to be presented in a forthcoming study, will provide a crucial test for the presence of a viscously heated disk spectrum, thus offering validation of the FUor-like classification proposed here.

\section{Conclusions}
\label{sect:conclusion}
We report the discovery of a large-amplitude infrared outburst from the deeply embedded Class I protostar NGC~7538~MIR, on the shell swept up by the \ion{H}{ii} region of NGC~7538 at a distance of $\sim$2.7~kpc. The source shows a rapid rise followed by a prolonged high-luminosity state lasting more than five years, with a total amplitude of $\Delta K_s \simeq 5$~mag, consistent with an FUor-type accretion outburst rather than a short-duration eruptive event. The long-term mid-IR light curves and color evolution trace a clear sequence of physical processes, transitioning from pre-outburst variability dominated by circumstellar extinction to accretion-driven luminosity enhancement near the outburst peak, and finally to post-outburst thermal and chemical relaxation of the circumstellar environment, likely regulated by water-ice sublimation and re-condensation. NGC~7538~MIR represents one of the most distant FUor-type eruptive YSOs identified to date, demonstrating that FUor-type episodic accretion operates not only in nearby low-mass star-forming regions, but also in heavily obscured, clustered environments associated with intermediate to high-mass star formation.

\begin{acknowledgements}
The authors acknowledge the in-depth reviews of anonymous referees. This work is supported by the National Natural Science Foundation of China (NSFC, grant No. 12373030). We would like to thank the CFHT Operations, Instrumentation, and Software Groups for their contributions and diligence in maintaining observatory operations; the CFHT Astronomy Group for their observation coordination and data acquisition efforts; and the CFHT Finance \& Administration Group for their contributions to the management and administration of the observatory. This work makes use of data products from the Near-Earth Object Wide-field Infrared Survey Explorer (NEOWISE), which is a joint project of the Jet Propulsion Laboratory/California Institute of Technology and the University of Arizona, and is funded by the National Aeronautics and Space Administration. 
\end{acknowledgements}




\label{lastpage}

\bibliographystyle{raa}
\bibliography{refs}
\end{CJK*}
\end{document}